\documentclass[%
 aip,
 amsmath,amssymb,
 reprint,%
]{revtex4-1}

\usepackage{graphicx}
\usepackage{dcolumn}
\usepackage{bm}

\usepackage[utf8]{inputenc}
\usepackage[T1]{fontenc}
\usepackage{mathptmx}

\begin{document}

\preprint{AIP/123-QED}

\title[]{Electromechanics in vertically coupled nanomembranes}

\author{Sepideh Naserbakht, Andreas Naesby, and Aur\'{e}lien Dantan}
\email{dantan@phys.au.dk}
\affiliation{Department of Physics and Astronomy, University of Aarhus, DK-8000 Aarhus C, Denmark}

\date{\today}

\begin{abstract}
We investigate the electromechanical actuation of a pair of suspended silicon nitride membranes forming a monolithic optomechanical array. By controlling the membrane resonators' tensile stress via a piezoelectrically controlled compressive force applied to the membrane chip we demonstrate noninvasive tuning of their mechanical mode spectrum, as well as strong intermode electromechanical coupling. Piezoelectric actuation is also shown to enhance the nonlinear response of the membranes, which is evidenced either by parametric amplification of the fundamental  mode thermal fluctuations or by resonant driving of these modes into high amplitude states. Such an electro-optomechanical membrane array represents an attractive tunable and versatile platform for sensing, photonics and optomechanics applications.
\end{abstract}

\maketitle

{\it Introduction}--Electro-optomechanical systems involving high-quality nano/micromechanical resonators and integrating electric and optical degrees of freedom~\cite{Midolo2018} are widely studied for sensing and photonic applications~\cite{Ekinci2005,Li2007}, as well as for fundamental investigations of the effects of radiation pressure in the context of optomechanics~\cite{Aspelmeyer2014}. 
Engineering linear and nonlinear electromechanical couplings in monolithic nanoresonator arrays is interesting for a wide range of ultrasensitive measurements~\cite{Ekinci2005}, photonics~\cite{Fan2019}, as well as for investigating collective dynamics, such as parametric resonances~\cite{Turner1998}, synchronization~\cite{Shim2007,Matheny2014} or coherent phonon manipulations~\cite{Mahboob2012,Faust2013,Okamoto2013,Mahboob2014,Yang2019,Huber2019}.

The combination of small effective mass, large area and high mechanical frequencies/quality factors makes suspended membranes made of low-loss material such as silicon nitride excellent resonators for optical sensing~\cite{Reinhardt2016,Guo2017,Naesby2017} or electrooptical conversion~\cite{Bagci2014,Andrews2014,Fink2016,Haghighi2018} applications. The optomechanical interaction of such a membrane with an optical cavity field~\cite{Thompson2008,Wilson2009} also allows for cooling its vibrations down into the quantum regime and observing radiation pressure-induced quantum effects~\cite{PurdySCI2013,PurdyPRX2013,Underwood2015,Peterson2016,Nielsen2017,Rossi2018}. 

Placing arrays of nanomembranes into an optical cavity~\cite{Piergentili2018,Gartner2018,Wei2019} furthermore opens up for exciting investigations of strong coupling and collective optomechanics~\cite{Bhattacharya2008,Hartmann2008,Xuereb2012,Seok2012,Xuereb2014,Kipf2014,Nair2016}, as enhanced optomechanical effects are predicted, effective phonon-phonon interactions can be engineered and phenomena involving multiple electromagnetic modes and resonators can be studied. Tunability of the individual mechanical elements in such optomechanical arrays and integration of the electric degree of freedom are highly desirable, if not essential, for investigating collective effects such as collectively enhanced radiation pressure forces~\cite{Xuereb2012,Nair2016}, superradiance~\cite{Kipf2014}, synchronization~\cite{Bemani2017}, topological interactions~\cite{Xu2016}, coherent phonon dynamics~\cite{Xuereb2014} or entanglement and multimode squeezing generation~\cite{Bhattacharya2008,Hartmann2008,Patil2015,Pontin2016}.

In this letter we investigate the electromechanical actuation via piezoelectricity of a pair of suspended silicon nitride membranes forming a monolithic optomechanical array. The application of a piezoelectric compressive force to one of the membrane chips allows for modifying the tensile stress of the membranes and, thereby, for tuning their mechanical mode spectrum without deteriorating the mechanical quality factors of the resonances. Such a scheme was recently applied to a single membrane resonator~\cite{Wu2018} and to a pair of distant membranes in an optical cavity~\cite{Wei2019}. We demonstrate here that the vibrational mode frequencies of two membranes in an 8.5 $\mu$m-long monolithic array can be tuned to degeneracy and strongly coupled via the application of a static bias voltage to the piezoelectric transducer. We also observe parametric amplification~\cite{Rugar1991} of the thermal fluctuations of their fundamental modes and demonstrate an enhancement of the nonlinear response of both membranes, evidenced by a lowering of the parametric oscillation thresholds and whose origin we discuss for each resonator.

\begin{figure}
\includegraphics[width=\columnwidth]{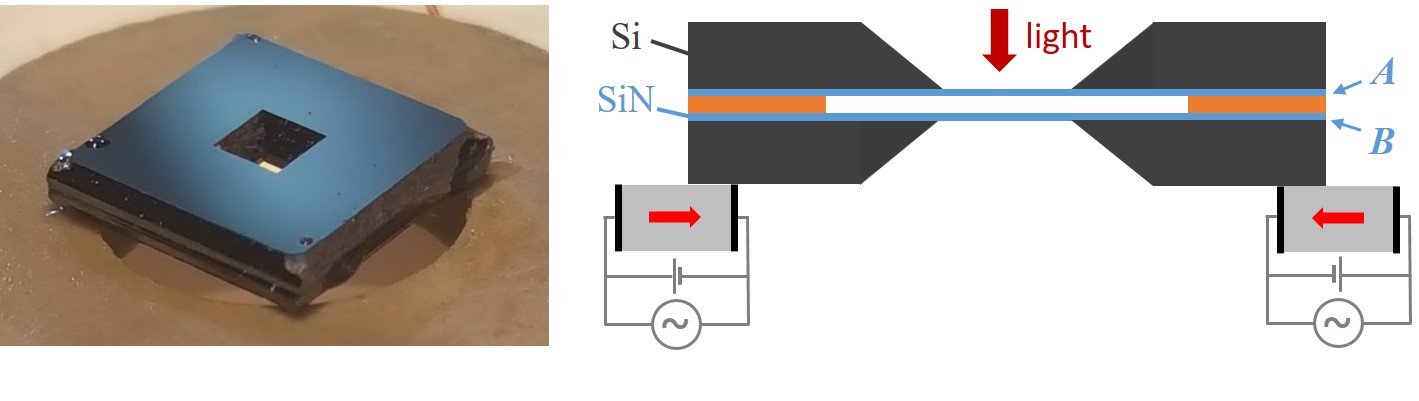}
\caption{Left: photograph of the array mounted on a ring piezoelectric transducer. Right: schematic transverse cut of the assembly (not to scale). The horizontal red arrows indicate the direction of the compressive piezoelectric force for a positive applied dc-voltage.}
\label{fig:array}
\end{figure}

{\it Electro-optomechanical array}--The array used in this work consists in a pair of commercial~\cite{Norcada}, high-stress, 500 $\mu$m-square stochiometric silicon nitride thin films (thickness 92 $\mu$m) deposited on a 5 mm-square silicon chip (thickness 500 $\mu$m). The chips were assembled parallel with each other with an 8.5 $\mu$m intermembrane separation (multilayer Si/Si/SiN spacer in Fig.~\ref{fig:array}) and their sides glued together at the corners following the method of Ref.~\cite{Nair2017}. Three corners of one of the chips were then glued to a 6 mm inner diameter piezoelectric ring transducer~\cite{Noliac}, as shown in Fig.~\ref{fig:array}. The vibrations of the membranes in vacuum ($10^{-7}$ mbar) are monitored by optical interferometry by measuring the transmission through the array of monochromatic light provided by a tunable external cavity diode laser (890-940 nm). The array then acts as a short, low-finesse Fabry-Perot cavity, whose length fluctuations can be analyzed by tuning the laser wavelength so as to maximize their amplitude, and by analyzing their frequency content using a low resolution bandwidth spectrum analyzer. This specific array exhibits a worse degree of parallelism after assembly (the tilt between the membranes is estimated to be around 2 mrad) than previous similar arrays~\cite{Nair2017}, resulting in a slightly reduced interferometric displacement sensitivity. The lowest square drummodes of the membranes of this array, with $\sim$ MHz frequencies and mechanical quality factors in the $10^5$ range, can still be reliably characterized, as done in~\cite{Naesby2017}. 

\begin{figure}
\includegraphics[width=0.85\columnwidth]{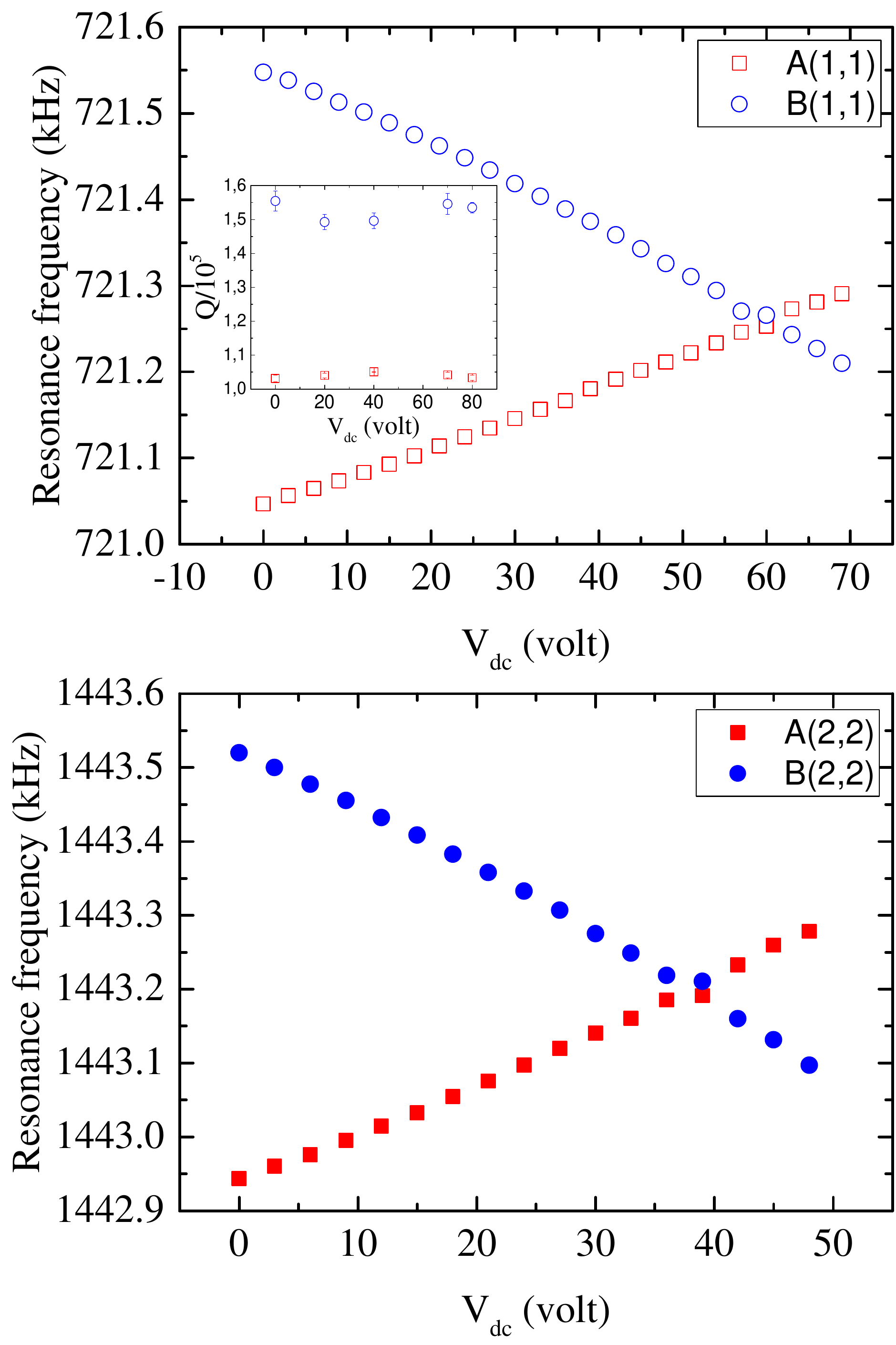}
\caption{(a) Resonance frequencies of the fundamental modes of both membranes as a function of the applied bias voltage $V_{\textrm{dc}}$. Inset: mechanical quality factors $Q_\alpha=\omega_\alpha/\gamma_\alpha$ ($\alpha=A,B$) versus $V_{\textrm{dc}}$. (b) Resonance frequencies of the (2,2) modes as a function of $V_{\textrm{dc}}$.}
\label{fig:actuation}
\end{figure}

{\it Tuning of mechanical mode spectrum}--In the tensile stress-dominated regime the vibrational mode frequencies of the membrane resonators are given by $\omega_{m,n}=\sqrt{\frac{\mathcal{T}}{\rho}}\frac{\pi}{a}\sqrt{m^2+n^2}$, where $\mathcal{T}=715$ MPa is the tensile stress, $\rho=2700$ kg/m$^3$ the density of silicon nitride, $a=500$ $\mu$m the lateral dimension of the membrane and $n$ and $m$ are strictly positive integers. When a static compressive force is applied to the silicon frame, the tensile stress of both films is modified in such a way that opposite, linear frequency shifts with similar magnitude are observed for the $(1,1)$ and $(2,2)$ modes of each membrane over the bias voltage range 0-80V,  as illustrated in Fig.~\ref{fig:actuation}. The compressive force on the bottom chip results in a reduced tensile stress for the silicon nitride film ($B$) deposited on this chip, and thereby a decreasing resonance frequency with positive bias voltage, as observed in~\cite{Wu2018}. The top chip, glued at its corners to the bottom one, then experiences an increased tensile stress and the resonance frequency of the $A$ membrane increases with bias voltage. Consistently with this picture, we tested that flipping the array and having the $A$ membrane chip glued to the transducer shows the opposite behavior, namely, a decrease in $A$ membrane's resonance frequencies and an increase in the $B$ membrane's. Piezoelectric biasing in the geometry of Fig.~\ref{fig:array} thus allows for achieving frequency degeneracy of the (1,1) and (2,2) for bias voltages $V_{\textrm{dc}}$ of 56 V and 38 V, respectively.  Futhermore, no noticeable effect of the bias voltage on the mechanical quality factors was observed (inset of Fig.~\ref{fig:actuation}(a)), demonstrating the noninvasive nature of the scheme.

\begin{figure}
\includegraphics[width=\columnwidth]{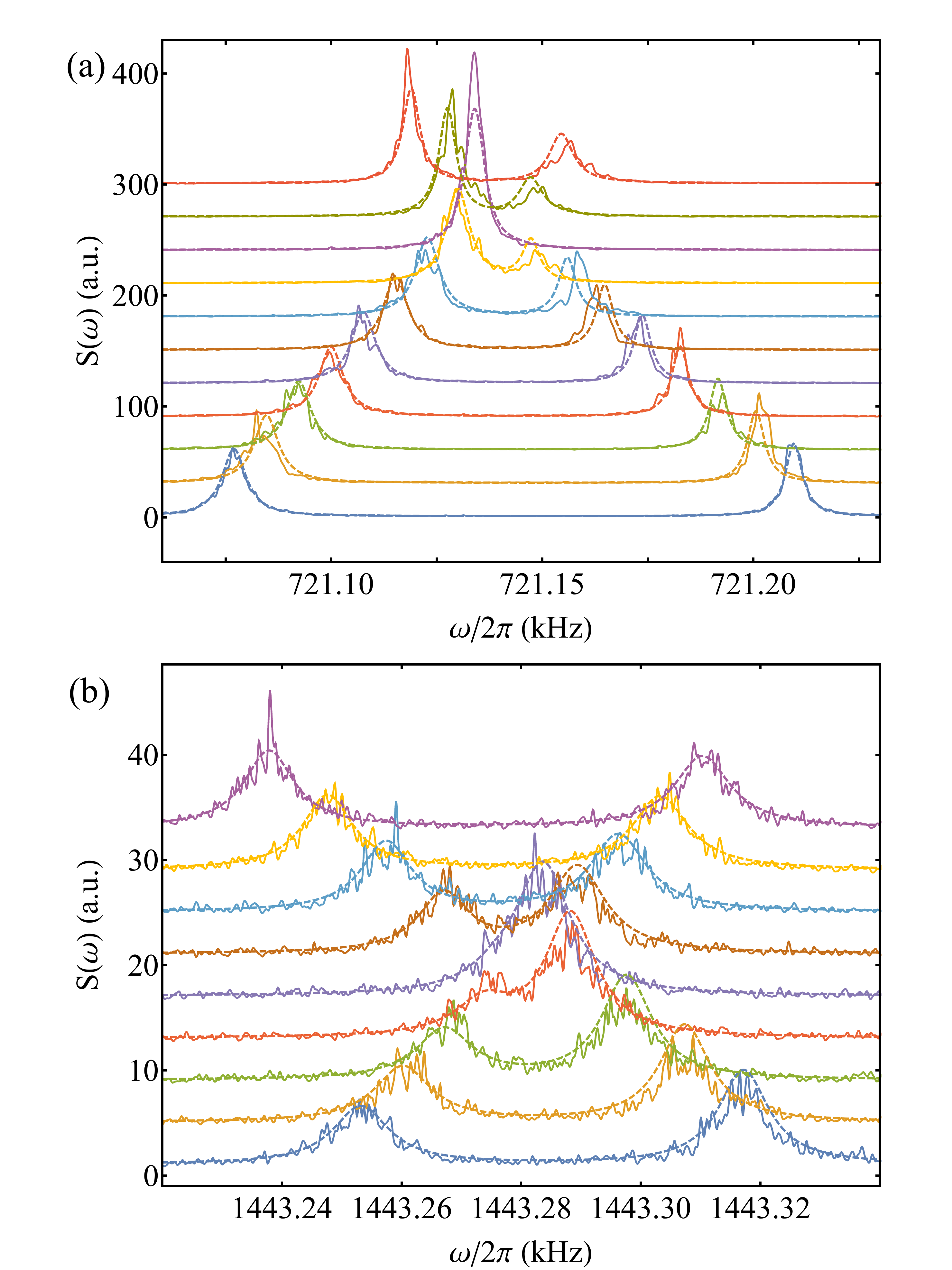}
\caption{Thermal noise spectrum for different bias voltages around the degeneracy point for the (1,1) modes (a) and (2,2) modes (b). $V_{\textrm{dc}}$ is varied from 40 V (lower curve) to 60 V (upper curve) in (a) and from 33 V (lower curve) to 41 V (upper curve) in (b). The spectra are offset vertically for clarity. The dashed lines show the results of a global fit to the coupled oscillator model described in the text.}
\label{fig:thermal}
\end{figure}

Interestingly, the intermode coupling via the frame/spacer structure can be investigated by analyzing the observed thermal noise spectrum around the degeneracy points, as shown in Fig.~\ref{fig:thermal}. This spectrum can be understood on the basis of a simple coupled oscillator model in which the dynamics of the mode amplitudes $x_{A,B}$ are given by
\begin{align}
\ddot{x}_A+\gamma_A\dot{x}_A+(\omega_A+\epsilon_AV_{\textrm{dc}})^2x_A&=\eta(x_B-x_A)+F_A,\\
\ddot{x}_B+\gamma_B\dot{x}_B+(\omega_B-\epsilon_BV_{\textrm{dc}})^2x_B&=\eta(x_A-x_B)+F_B,
\end{align}
where $\gamma_{A,B}$ are the mode mechanical damping rates, $\omega_{A,B}$ their resonance frequencies at zero bias voltage, $\epsilon_{A,B}V_{\textrm{dc}}$ the linear voltage dependent frequency shifts, $\eta$ the intermode coupling constant and $F_{A,B}$ the thermal noise forces (divided by the mode effective mass). Fourier transforming these equations readily yields the Fourier component amplitudes at frequency $\omega$, e.g.,
\begin{equation}
x_A(\omega)=\frac{\chi_B(\omega)F_A(\omega)+\eta F_B(\omega)}{\chi_A(\omega)\chi_B(\omega)-\eta^2}
\end{equation}
 where $\chi_{\alpha}(\omega)=(\omega_\alpha+\epsilon_\alpha V_{\textrm{dc}})^2+\eta-\omega^2-i\gamma_\alpha\omega$ ($\alpha=A,B$) and with a similar expression for $x_B(\omega)$ when exchanging subscripts $A$ and $B$. The interferometric signal measured by the spectrum analyzer is proportional to the noise spectrum of $x_A-x_B$, $S(\omega)$, whose analytical expression can then be obtained from the previous relations. The dashed lines in Fig.~\ref{fig:thermal} show the results of a global fit of the data to the model, fixing the mechanical damping rates and thermal force amplitudes using the spectra far from the degeneracy points, and leaving as free parameters the voltage-dependent frequency shift rates $\epsilon_{A,B}$, as well as the intermode coupling rate $\bar{\eta}$, defined by $\eta=\bar{\eta}(\omega_A+\omega_B)/2$. The resulting spectra are observed to match well the observed data, in particular in the avoided crossing region, where the intermode coupling plays a significant role. Remarkably, the extracted value for the (1,1) modes, $\bar{\eta}/(2\pi)\simeq 8$ Hz, is found to be slightly larger than both mechanical decay rates $\gamma_A/(2\pi)\simeq 7$ Hz and $\gamma_B/(2\pi)\simeq 4.5$ Hz, which places such an electromechanical array at the border of the strong coupling regime. For the (2,2) modes, whose mechanical quality factors are a bit lower, $\eta/(2\pi)\simeq 7$ Hz is slightly smaller than $\gamma_{A,B}/(2\pi)\simeq 10$ Hz. Such strong electromechanical couplings are promising for e.g. coherent phonon manipulations~\cite{Mahboob2012,Faust2013,Okamoto2013} or electro-optical conversion~\cite{Bagci2014,Andrews2014,Fink2016,Haghighi2018}.

\begin{figure}
\includegraphics[width=\columnwidth]{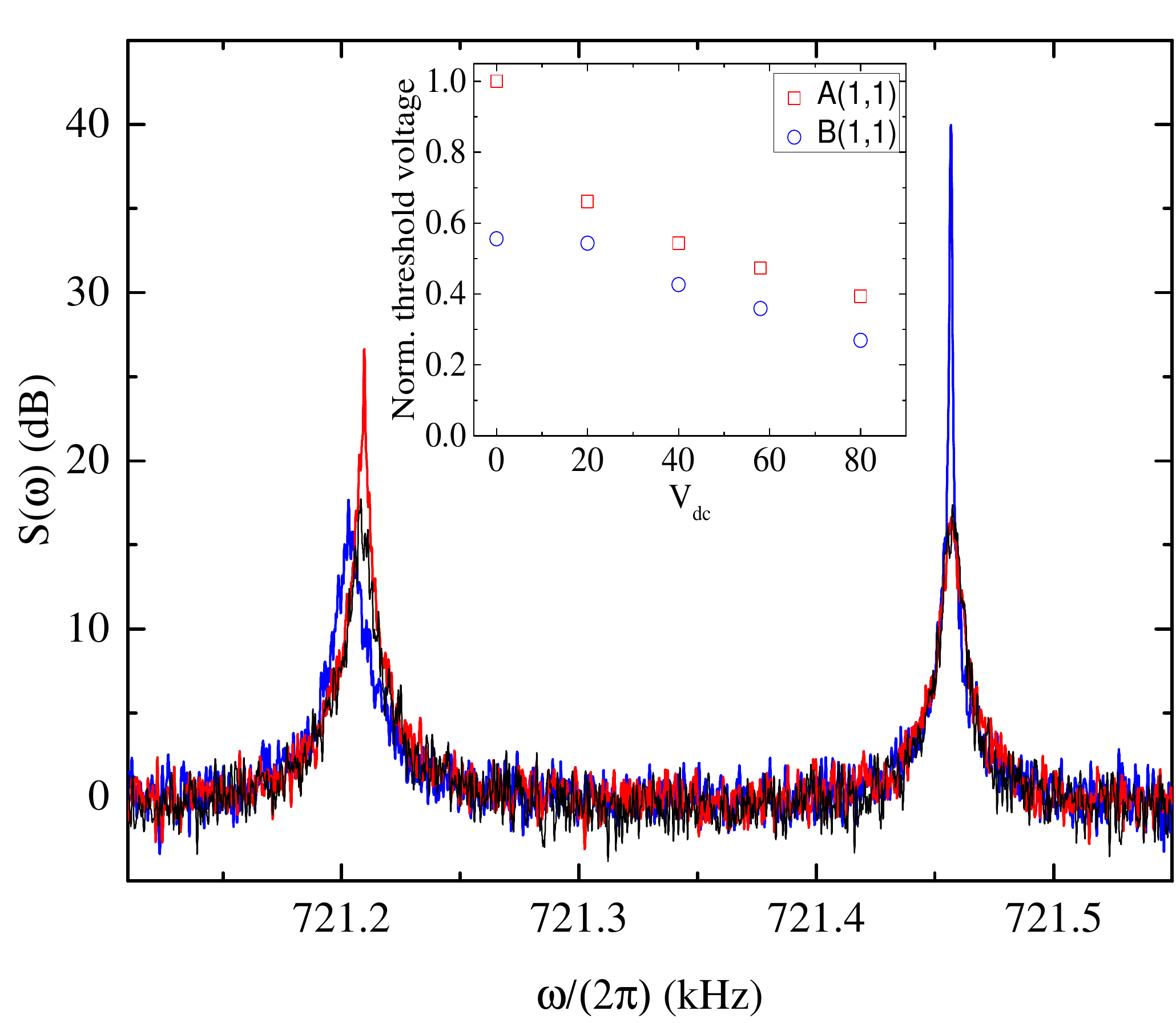}
\caption{Noise spectra of both fundamental modes when either the $A(1,1)$ mode (red, $V_{2\omega}=322$ mV) or the $B(1,1)$ mode (blue, $V_{2\omega}=300$ mV) is parametrically excited below the oscillation threshold and for $V_{\textrm{dc}}=40$ V. The black curve shows the thermal noise spectrum as a reference. Inset: parametric oscillation threshold voltage $V_\textrm{dc}^\textrm{th}$ for both fundamental modes (normalized to the 448 mV threshold voltage of the $A(1,1)$ mode at zero bias voltage) as a function of $V_{\textrm{dc}}$.}
\label{fig:parametric}
\end{figure}

{\it Parametric actuation}--We now turn to the piezoelectric tuning of the nonlinear response of the mechanics and investigate the parametric amplification~\cite{Rugar1991,Turner1998,Carr2000,Mahboob2008,Karabalin2010,Thomas2013,Mahboob2014,Seitner2017,Wu2018} of the thermal fluctuations of the fundamental modes of both membranes when, in addition to the dc bias voltage, a modulation at twice the mechanical resonance frequency, $V_{2\omega}\cos(2\omega t)$, is applied to the piezoelectric transducer. When one of these modes, with resonance frequency $\omega_\alpha$, is parametrically driven at $2\omega_\alpha$, its dynamics can be described by the Mathieu equation~\cite{Rugar1991}
\begin{equation}
\ddot{x}_\alpha+\gamma_\alpha\dot{x}_\alpha+\omega_\alpha^2[1+\zeta_\alpha\cos(2\omega_\alpha t)]x_\alpha=F_\alpha,
\label{eq:mathieu}
\end{equation}
where $\omega_\alpha$ includes the bias voltage shift and where $\zeta_\alpha$ is proportional to the parametric modulation amplitude $V_{2\omega}$. The bias voltage is chosen so as to operate away from the degeneracy point for the (1,1) modes, as well as far from resonance with the (2,2) modes, whose frequencies are close to the second harmonic of the fundamental frequencies (the absence of excitation of these modes was verified experimentally). Figure~\ref{fig:parametric} shows examples of parametrically amplified noise spectra of both modes at zero bias voltage. The fluctuations of the mode which is parametrically driven resonantly are observed to strongly increase while its noise spectrum becomes narrower, as the parametric modulation amplitude increases, until the parametric oscillation threshold is reached when $\zeta_\alpha\sim 2/Q_\alpha$. Parametric gains of a few tens to a few hundreds are typically observed before the oscillation threshold is reached. The inset of Fig.~\ref{fig:parametric} shows the variations with the bias voltage of the parametric oscillation threshold voltage of both modes; application of the bias voltage strongly reduces the parametric oscillation threshold of both membrane modes, in a seemingly relatively similar fashion. 

\begin{figure}
\includegraphics[width=0.87\columnwidth]{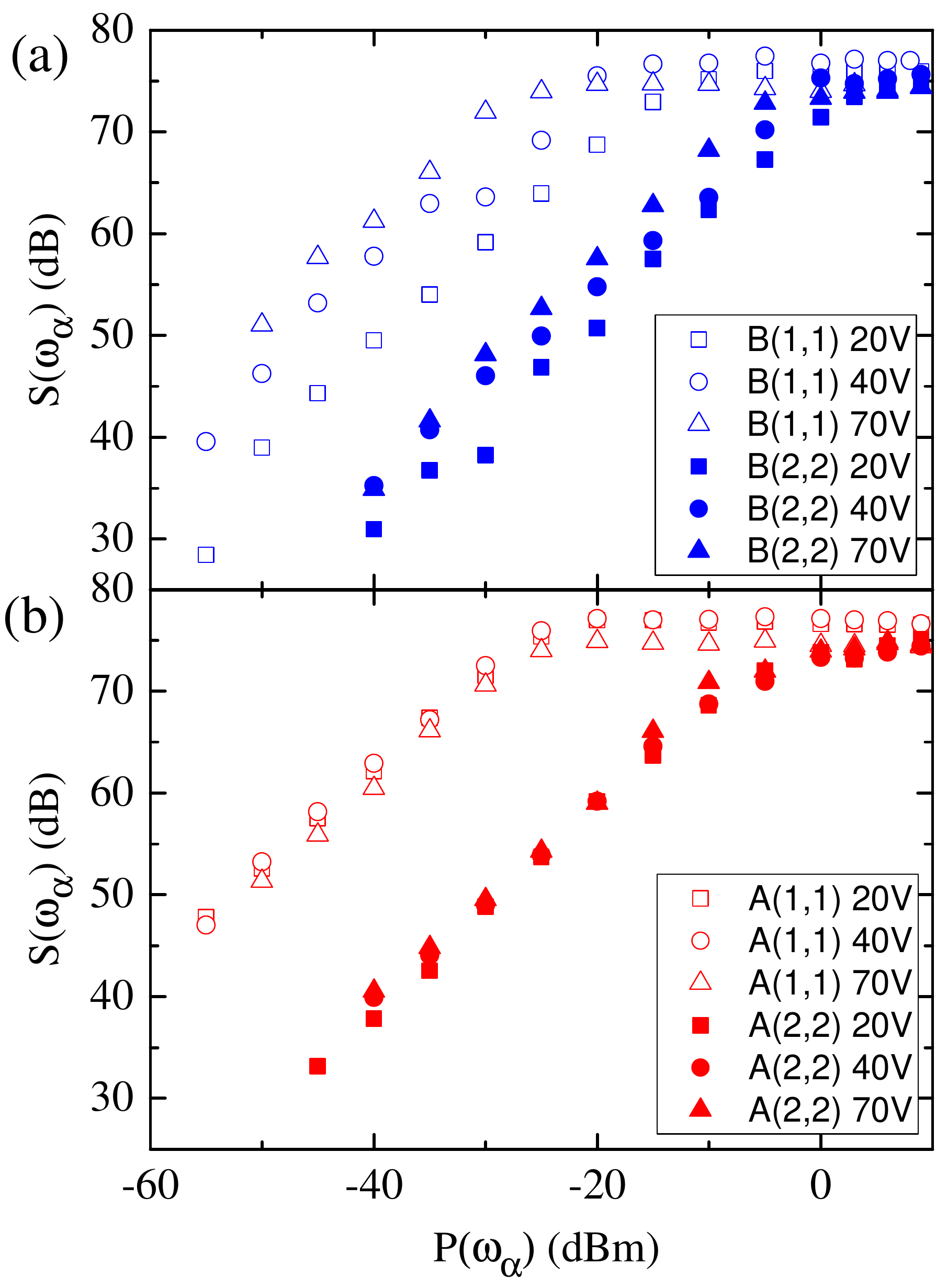}
\caption{Driven response of the (1,1) and (2,2) modes of membrane $B$ (a) and $A$ (b): Noise spectrum at resonance (dB) as a function of the power of the modulation voltage at the mechanical resonance frequency $\omega_\alpha$.}
\label{fig:drive}
\end{figure}

The effect of the bias voltage on the response of each membrane is quite different in nature, though. To assess the effect of biasing on the dynamical response of the membranes, the (1,1) and (2,2) modes of each membranes were driven independently at their mechanical resonance frequency, for a fixed bias voltage and increasing modulation amplitudes, as shown in Fig.~\ref{fig:drive}. A linear response is observed over a wide range of modulation amplitudes, before nonlinearities kick in~\cite{Yang2019}. The linear response of the modes of membrane $B$--which is directly coupled to the piezoelectrically stressed silicon frame--is strongly affected by the bias voltage; in particular, its increased response at $\omega_{2,2}\simeq2\omega_{1,1}$ accounts well for the decrease of the parametric oscillation threshold with $V_{\textrm{dc}}$. 

\begin{figure}
\includegraphics[width=0.85\columnwidth]{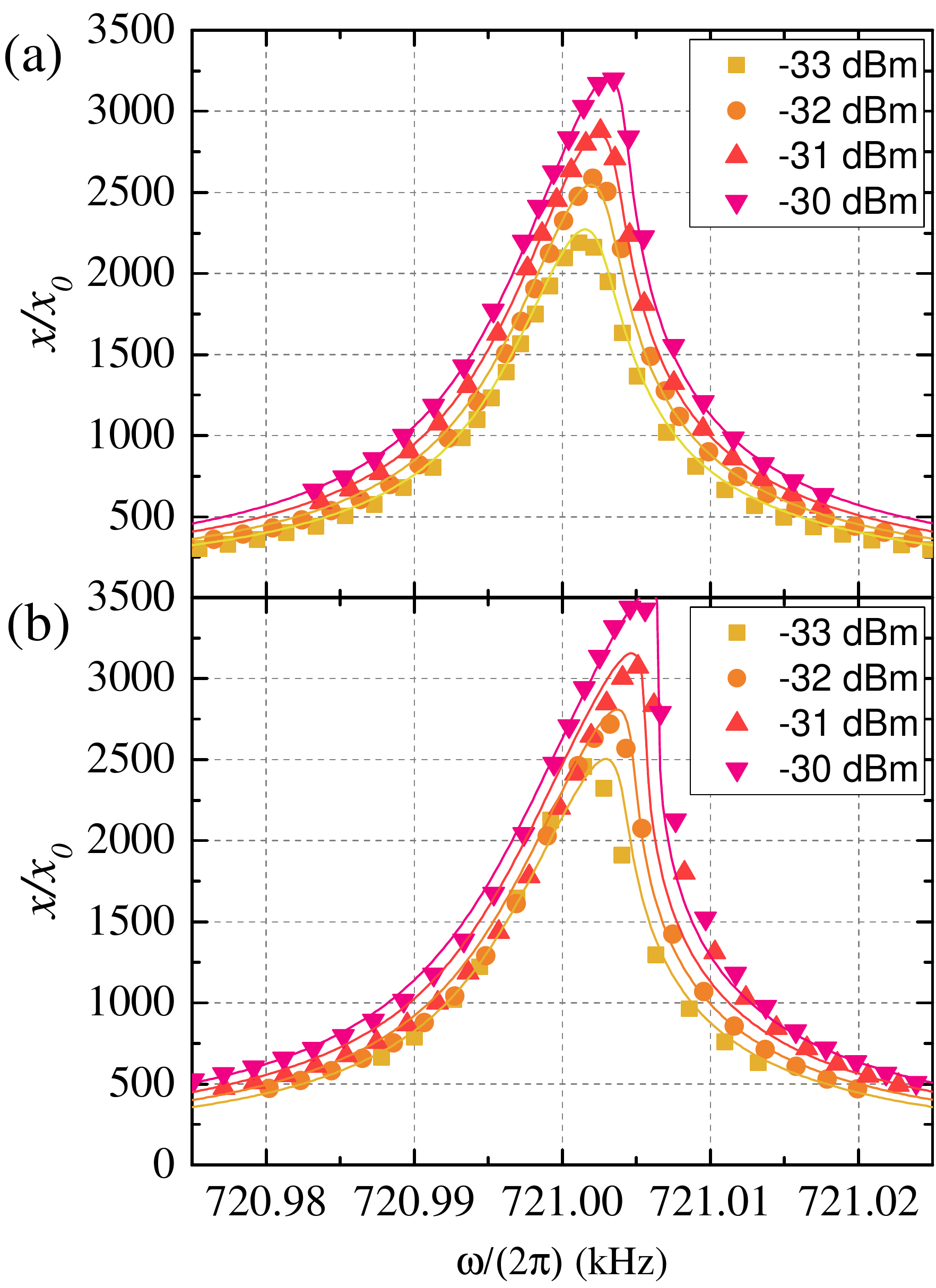}
\caption{Amplitude (normalized to the thermal motion amplitude $x_0$) of the $A(1,1)$ mode as a function of drive frequency around $\omega_{1,1}$, for different drive powers (a power of -30 dBm corresponds to an applied modulation voltage amplitude of 9.2 $\mu$V): (a) $V_{\textrm{dc}}=20$ V and (b) $V_{\textrm{dc}}=80$ V. The solid lines show the results of global fits to the nonlinear Duffing oscillator model discussed in the text.}
\label{fig:duffing}
\end{figure}

In contrast, such a direct parametric modulation of the spring constant does not explain the lowering of the parametric oscillation threshold for $A(1,1)$, as the linear response of the modes of membrane $A$--whether at the fundamental or the second harmonic frequency--is fairly independent of the bias voltage. To investigate the role of the biasing on the nonlinear response of membrane $A$ further, its fundamental mode was driven in the high amplitude regime before the onset of bistability. Frequency scans around $\omega_{1,1}$ of the signal measured by the spectrum analyzer in zero-span mode were performed for different drive powers and different bias voltages. These scans, shown in Figs.~\ref{fig:duffing}, clearly show nonlinearly distorted resonance profiles. Such profiles can be accurately reproduced by introducing a cubic Duffing nonlinearity in the equation of motion for the mode dynamics
\begin{equation}
\ddot{x}_\alpha+\gamma_\alpha\dot{x}_\alpha+\omega_\alpha^2[1+\beta_\alpha x_\alpha^2]x_\alpha=F_\alpha+F_\omega\cos(\omega t),
\label{eq:duffing}
\end{equation}
where $F_\omega$ is the amplitude of the driving force at frequency $\omega$ and $\beta$ is the Duffing nonlinearity coefficient. For a small nonlinearity and neglecting the thermal force, the Fourier component amplitude at $\omega$ is approximately given by a solution of the implicit equation~\cite{Jordan}
\begin{equation}
|x_\alpha(\omega)|=\frac{F_\omega}{\sqrt{(\omega_\alpha^2(1+\tfrac{3}{4}\beta_\alpha |x_\alpha(\omega)|^2)-\omega^2)^2+\gamma_\alpha^2\omega^2}}.
\end{equation}
The solid lines in Figs.~\ref{fig:duffing} show the results of global fits of this equation to the data, using low drive power scans to fix $\omega_\alpha$ and $\gamma_\alpha$ and leaving as free parameters $\beta_\alpha$ and a global amplitude for the driving force, the respective amplitudes being apppropriately scaled by the known applied powers. The fits match well the observed spectra and yield values of $\beta_\alpha$ (in units of $x_0^2\times 10^{-12}$) of $1.11\pm 0.16$ and $1.75\pm0.07$ for $V_{\textrm{dc}}=20$ V and 80 V, respectively. Application of a bias voltage thus increases the nonlinear response of membrane $A$ to a driving force at the resonance frequency $\omega_\alpha$. Since the nonlinear Duffing term in $x_\alpha^2$ in Eq.~(\ref{eq:duffing}) under such a driving can be seen to represent an effective modulation at twice the resonance frequency, as in Eq.~(\ref{eq:mathieu}), the bias voltage-dependent Duffing nonlinearity coefficient $\beta_\alpha$ is thus expected to be proportional to the parametric modulation coefficient $\zeta_\alpha$, itself inversely proportional to $V_\textrm{dc}^{\textrm{th}}$. This is corroborated by the $\sim$60\% increase in $\beta_\alpha$ when increasing the bias voltage from 20 V to 80 V, which matches well the observed reduction in the parametric oscillation threshold voltage by about the same amount. Piezoelectric actuation can thus be used to enhance the nonlinear response of both resonators, which is interesting for generating thermomechanical squeezing or entanglement~\cite{Rugar1991,Mahboob2014,Wu2018,Huber2019}, among others.

\textit{Conclusion}--A simple and noninvasive scheme, based on piezoelectrically-induced stress control, for tuning the vibrational mode frequencies and the nonlinear response of high-Q suspended membrane resonators in a monolithic optomechanical array has been demonstrated. While enhancing their nonlinear response is useful for various sensing applications, tuning the mechanics and engineering electromechanical couplings in such electro-optomechanical arrays is essential for future collective optomechanics investigations.

\begin{acknowledgments}
We acknowledge support from the Velux Foundations.
\end{acknowledgments}

\nocite{}
\bibliographystyle{apsrev4-1}
\bibliography{piezo_array_bib}

\end{document}